\documentclass[
]{ceurart}

\usepackage{xcolor}

\usepackage{amsmath}
\usepackage{pdfpages}
\usepackage{verbatim}
\usepackage{hyperref}
\usepackage{combelow}
\usepackage{graphicx}
\usepackage{todonotes}

\usepackage{caption}
\usepackage{subcaption}

\begin{document}

\copyrightyear{2021}
\copyrightclause{Copyright for this paper by its authors.
  Use permitted under Creative Commons License Attribution 4.0
  International (CC BY 4.0).}

\conference{$7^{th}$ International Workshop on Satisfiability Checking and Symbolic Computation, 12 Aug 2022, Haifa Israel}

\title{SMT-Solving Induction Proofs of Inequalities}


\author[1,2]{Ali K. Uncu}[%
orcid=0000-0001-5631-6424,
email=aku21@bath.ac.uk,
url=http://akuncu.com,
]
\address[1]{Austrian Academy of Science, Johann Radon Institute for Computational and Applied Mathematics, Linz, Austria}

\author[2]{James H. Davenport}[%
orcid=0000-0002-3982-7545,
email=masjhd@bath.ac.uk,
url=https://people.bath.ac.uk/masjhd,
]
\address[2]{University of Bath, Faculty of Science, Department of Computer Science, Bath, UK}

\author[3]{Matthew England}[%
orcid=0000-0001-5729-3420,
email=Matthew.England@coventry.ac.uk,
url=https://matthewengland.coventry.domains,
]
\address[3]{Coventry University, Research Centre for Computational Science and Mathematical Modelling, Coventry, UK}

\begin{abstract}
This paper accompanies a new dataset of non-linear real arithmetic problems for the SMT-LIB benchmark collection. The problems come from an automated proof procedure of Gerhold--Kauers, which is well suited for solution by SMT. The problems of this type have not been tackled by SMT-solvers before.  We describe the proof technique and give one new such proof to illustrate it. We then describe the dataset and the results of benchmarking. The benchmarks on the new dataset are quite different to the existing ones.  The benchmarking also brings forward some interesting debate on the use/inclusion of rational functions and algebraic numbers in the SMT-LIB.
\end{abstract}


\maketitle

\section{Introduction}
\label{Sec1}

Satisfiability Modulo Theories (SMT) fuses powerful modern SAT solvers with software from specialised theory domains to tackle satisfiability problems where the logical atoms are statements in that domain.  The SMT-LIB \cite{SMTLIB} defines a common language for SMT-solvers to use and maintains a set of benchmarks organised according to the various theory domains.   

In many cases, the algorithms for those domains have been traditionally implemented in computer algebra systems (although as described in \cite{AAB+16a}, such algorithms require adaptation before they can be used efficiently within SMT).  There is continuing progress in algorithms for such domains, driven in part by the connections built between symbolic computation and the satisfiability checking communities, by the SC-Square project \cite{AAB+16a} and others.  

One of the SMT theory domains most closely aligned with symbolic computation, and the domain we consider, is \verb+QF_NRA+.  In this case the solver seeks to answer a question on the existence of real variables $x_1, \dots, x_k$ to solve a logical formula in which each atom is a (potentially non-linear) polynomial constraint.  There is a significant number of benchmarks in the SMT-LIB for this domain, however, there are relatively few sources of these examples, and vast majority come from a single theorem-proving application.  In this paper we describe a new collection of examples which we have contributed, originating from inductive proof of some inequalities.  We seek to (a) broaden the \verb+QF_NRA+ benchmark set to allow for better development of solvers; and (b) encourage further additions from other new application domains by demonstrating how well solvers can do on such problems.

\subsection{SMT for QF\_NRA}

In the \verb+QF_NRA+ domain solvers tackle satisfiability problems whose atoms are of the form $p \sigma 0$ where $p := p( x_1, x_2, \dots, x_k) \in \mathbb{Q}[x_1,x_2,\dots,x_k]$ is a polynomial in variables $x_1,\dots ,x_k$ with rational coefficients, and $\sigma \in\{>,<,\geq, \leq, =, \not =\}$.  

Such problems are usually tackled in the Lazy SMT paradigm where a SAT-solver proposes solutions to the logical structure which are then checked for validity in the theory domain: deciding whether the corresponding set of polynomials constraints can be satisfied together.  In other words, we check the following, where $p_i$ and $\sigma_i$ are defined as above:
\begin{equation}
\label{eq:exists_statement} 
\exists x_1,x_2,\dots, x_k\ (p_1 \sigma_1 0 \land p_2\sigma_2 0\land p_n \sigma_n 0).
\end{equation}
Note that this is fully conjunctive and will involve only a subset of the atoms in the original formula.  
The answer will either be a set of assignments for the variables $(x_1, \dots, x_k)$ to \emph{witness} the existence, or a confirmation that this is \emph{unsatisfiable}. The unsatisfiable confirmation confirms that there is no single point in $\mathbb{R}^k$ space that could satisfy the relations.

There are many ways of tackling this conjunction.  One expensive but well established method is to calculate the Cylindrical Algebraic Decomposition (CAD) \cite{Collins1975} of the variable space to be sign-invariant for the polynomials $p_i$, and then check the regions for existence of such a point.  CAD was developed for the more general problem of Quantifier Elimination (QE) over the reals.  Such a heavy procedure is clearly far more work than required, at least when the problem is satisfiable and we need find only a single point.  An adaptation of CAD for this purpose was presented in \cite{KA20} to allow for early termination and repeated calls.

Another approach is to re-purpose the theory of CAD so that it better aligns to the satisfiability methodology, of searching for a model and learning from conflict.  This is the approach taken in the Cylindrical Algebraic Covering method of \cite{ADEK21} and the NLSAT algorithm of \cite{JdM12}.  Both build model solutions gradually variable by variable:  the former learns by identifying that a model cannot be extended and using CAD projection to rule out an interval around the top dimension of the current model; the latter learns by building an entire CAD cell whose description defines new atoms of the logical formula.

In addition to these CAD based methods there is a variety of incomplete methods implemented which may will not always give a solution, but when they do are often far faster, e.g. Virtual Term Substitution \cite{Sturm2018} or incremental linearization \cite{CGIRS18c}.  Although these techniques all root from formal arguments, there is not a clear answer as to which method is best in general or on a particular instance, and so there is a need for careful benchmarking. Also, different problems sets have their own \emph{flavour}, and so benchmarking on them individually is valuable too.  This set, rooting from mathematical inequalities, is internally diverse. Although the number of the variables do not go beyond 6 at any time, the natural appearance of rational functions in some problems and the degree of a single variable spiking in some other problems can be considered as some different characteristics within this sample set.

\subsection{Problems from inequalities}

We will discuss a family of \verb+QF_NRA+ problems new to the SMT-LIB, and to the best of the authors' knowledge, not tackled before using SMT.

In 2005, Gerhold and Kauers presented an algorithm that attempts induction proofs with great success \cite{GK05}. Their original formulation, and later Kauers' \texttt{ProveInequality} function in the Mathematica package \texttt{SumCracker} \cite{Kauers2006}, uses CAD to make these proofs. This method and the implementation has been used successfully applied in many works to automatically prove combinatorics and special function related inequalities \cite{GK05, Kauers2006, Kauers2005, Dixitetal2016a, KauersPillwein2010a, KP07, GK06, Kauers2011}. These applications utilised computer algebra, but the underlying algorithm is actually asking a sequence of satisfiability questions that terminates with a positive answer if it can be shown that a logical structure of the form \eqref{eq:exists_statement} is unsatisfiable. 

Later in the paper we will sketch the main ideas behind this procedure by proving the following result.  Let $k$ and $n$ be positive integers, and let $x_1,x_2,\dots x_n \in \mathbb{R}^{+}$. Then if $x_1+x_2+\dots+x_n =n$, we have that 
\begin{equation}
\label{eq:example} 
\sum_{i=1}^n x_i^k \, \frac{x_i^4+x_i^2+1}{x_i^2+x_i+1} \geq n \prod_{i=1}^n x_i.
\end{equation} 
This problem appeared as a generalization of a Monthly Problem in the American Mathematical Monthly  \cite{AmerMathvol125GenProb}. Until now the inequality \eqref{eq:example} only had a human proof. That proof required using an inequality on carrying positive exponents inside a finite sum, followed by an inequality of Chebyshev, and then an inequality between arithmetic-geometric means. However, by following the Gerhold--Kauers method, we will prove a stronger version of this inequality without any prior knowledge and with minimal human interference.

\subsection{New dataset}

We have put together a dataset of 300 problems in the SMT-LIB language, derived from the application of the Gerhold--Kauers method to examples given in \cite{GK05, Kauers2006, Kauers2005,  KauersPillwein2010a} and \eqref{eq:example}.  These have been submitted for inclusion in the 2022 release of the SMT-LIB. 

Unlike other problem sets in the SMT-LIB, a quarter of these problems have constraints that involve rational functions instead of purely polynomial constraints. This new characteristic also calls into question how best to pre-process such objects.  There are at least two ways of clearing any non-constant denominators to get an equivalent expression with polynomial constraints, and then there is the handling of zeros of any denominators. For every problem involving a rational function, we generated two equivalent problems in polynomials, where we handled the denominators in a different way.  We will observe different solver behaviour depending on the conversion method used.

\subsection{Plan of the paper}

The organization of this paper is as follows. In Section~\ref{Sec2}, we will briefly introduce the Gerhold--Kauers method by using it on the new example. Then in Section~\ref{Sec3} we will discuss different ways of clearing denominators before presenting our dataset and some benchmarking results for it in Sections~\ref{Sec4} and \ref{Sec5}. 
We finish with some conclusions.

\section{The Gerhold--Kauers method to use CAD for Induction Proofs}
\label{Sec2}

\subsection{General idea}

A mathematical induction proof --- at its core --- is a finite set of initial conditions together with the logical structure of the problem implying the correctness of the next step. We require a discrete parameter, say $n$, for the indexing of the initial conditions. Let, our general claim,  $\phi$ be a logical formula (or a collection of logical formulae) in $n$ (without loss of generality $n\in\mathbb{Z}^+$) and possibly other variables. We would like to prove the correctness of $\phi$ by complete induction of $n$. The construction of $\phi$ is done through a difference field construction: we will brush over that and invite any interested readers to visit Gerhold and Kauers' original paper \cite{GK05}. 

It is not clear from which starting point and with how many (if any) initial conditions one can gather satisfactory knowledge to prove the induction step. Hence one needs to start with the selection of a $t$ and $r$ (both being most likely 1) and attempt to show 
\begin{equation}
\label{eq:basicInduction} 
\psi\land (\phi\land s(\phi) \land \dots\land s^{r-1}(\phi)) \Rightarrow s^{r}(\phi),
\end{equation} 
where $\psi$ is a conjunction of all known assumptions on the parameters, and $s^k(\phi)$ is the $k$-th shift (in $n$) of the original statement $\phi$. Let $[\phi]_k$ be the explicit evaluation of $\phi_n$ at the instance $k\in\mathbb{Z}_{\geq 0}$. If we can also confirm that each initial condition $[\phi]_k$ for $k=t,\dots, t+r-1$ holds together with \eqref{eq:basicInduction} then we get an induction proof for all $n\geq t$. 

In their paper \cite{GK05}, Gerhold and Kauers decide to attempt refuting \eqref{eq:basicInduction} by instead attempting to deduce that 
\begin{equation}
\label{eq:neg_induction}
\psi\land (\phi\land s(\phi) \land \dots\land s^{r-1}(\phi))\land \lnot s^{r}(\phi)
\end{equation} 
would be false. Moreover, they do it in an efficient and iterative way by checking $[\phi]_n$'s at each step and extending $r$ if \eqref{eq:neg_induction} is still satisfiable for some selection of variables. A possible variable selection might be far away from the original problem, however, such an instance triggers the algorithm to iterate (pick a larger $r$) and repeat the process.

\subsection{A New Proof of (2)}

As an initial step towards the proof of \eqref{eq:example}, let us start with the claim that, for $x_i>0, i=1,\dots, n$, if $\sum_{i=1}^n x_i = n$, then
\begin{equation}
\label{eq:1}
\prod_{i=1}^n\ x_i \leq 1.
\end{equation} 
The case $n=1$ is obvious. The difference ring construction would define x in the place of $x_n$. Then we define another three variables $X$, $Y$, and $Z$ and their shifts in $n$: $s(X) = X + 1$, $s(Y) = Y + s(x)$, and $s(Z) = Z s(x)$, where  $s(\cdot)$ is the shift of the variable inside (the next element in the sequence) and $s(x)$ is kept as a new variable added to the problem. Here $X$ simulates $n$, $Y$ simulates the sum $\sum_{i=1}^n x_i$ and $Z$ simulates the product $\prod_{i=1}^n\ x_i$. Assuming $t=r=1$, the logical statement we are trying to refute is 
\begin{align} \label{eq:lf}
&\quad \big( x>0 \land s(x)>0 \land X=Y \land s(X)=s(Y) \big) \land (Z \leq 1) \land \lnot (s(Z) \leq 1)\\ 
\nonumber&= \big( x>0\land s(x)>0 \land X=Y \land X+1=Y+s(x) \big)\land (Z\leq 1) \land (Z s(x) > 1), 
\end{align} together with the initial condition check $[Z]_1 =1 \leq1$.
It is very easy to see that the first logical sentence implies $s(x)=1$ and that together with the last two clauses yields a contradiction.  Therefore, confirming the claim for the initial conditions $[X]_1, [Y]_1$ and $[Z]_1$, the induction step holds and our claim is true for generic $n\geq 1$.

Similarly, we can prove 
\begin{equation}\label{eq:2} 
\sum_{i=1}^n  \frac{x_i^4+x_i^2+1}{x_i^2+x_i+1} \geq n,
\end{equation} 
under the assumptions $x_i >0$ for $i=1,\dots, n$ and $\sum_{i=1}^n x_i = n$. 
To simulate the sum on the left-hand side of \eqref{eq:2} and its iterations, is given as $[\hat{Z}]_1=x_1^2 - x_1 + 1$ and $s(\hat{Z})= \hat{Z} + (s(x)^4 +s(x)^2+1)/(s(x)^2 +s(x)+1)$.
For the proof, one can put together the logical formula to be refuted, similar to \eqref{eq:lf}, with $X$, $Y$ and the new variable $\hat{Z}$ and easily show that to be contradiction. 


In the same vein, we can prove
\begin{equation}\label{eq:3}
\sum_{i=1}^n (x_i -1)  \frac{x_i^4+x_i^2+1}{x_i^2+x_i+1} \geq 0, 
\end{equation}  following similar steps as above with the new variable $\tilde{Z}$, where $s(\tilde{Z})= \tilde{Z}+(s(x)-1)(s(x)^4 +s(x)^2+1)/(s(x)^2 +s(x)+1)$.
 

The next step needed to prove \eqref{eq:example} is to show
\begin{equation}
\label{eq:4}
\sum_{i=1}^n x_i^{j-1} (x_i -1)  \frac{x_i^4+x_i^2+1}{x_i^2+x_i+1} \geq 0,
\end{equation} 
for any $j\geq 1$.  For any fixed positive integer $j$ this can be done with a logical solver for \verb+QF_NRA+. In this example the logical formula to evaluate becomes 
\[x>0\land s(x)>0 \land X=Y \land X+1=Y+s(x) \land \overline{Z}\geq 0 \land \overline{Z}> s(x)^j(s(x)^3 -2s(x)^2 +2s(x)+1),
\]
where $\overline{Z}$ simulates the sum on the left-hand side of \eqref{eq:4}. From the logical structure, we can deduce that $s(x)=1$ and later conclude that $(\overline{Z}\geq 0) \land (\overline{Z} < 0)$ would yield a contradiction and prove \eqref{eq:4}. However, this is only possible to achieve on a computer for explicitly chosen positive integers $j$. Otherwise, since the input would not be a collection of polynomials/rational functions, we could not apply CAD (or other QE methods).

This is where we need a human touch to prove \eqref{eq:4} for arbitrary $j$ using \eqref{eq:3}. 
The case of all $x_i=1$ is trivially true.
Otherwise, since $\sum_{i=1}^n x_i=n$, there exists at least one $a \in \{ 1,2,\dots, n\}$ such that $x_a > 1$ and at least one $b\in \{ 1,2,\dots, n\}$ such that $x_b<  1$. 
Let $A$ be the set of all such indices between 1 and $n$ such that $x_a>1$. Similarly, let $B$ be the set of all indices of all $x_b$ such that $0<x_b<1$. $A$ and $B$ are both finite sets since all the indices are chosen from 1 to $n$. For non-empty $A$ and $B$, notice that $x_a^{j-1}\geq x_a>1$ and $0< x_b^{j-1} \leq x_b$ for any $a\in A$ and $b\in B$. So by multiplying the $i$-th summand of \eqref{eq:3} with $x_i^{j-1}$, we either keep the summand the same (if $j=1$) or increase the contribution of the positive terms if $i\in A$. Similarly, by multiplying the $i$-th summand of \eqref{eq:3} with $x_i^{j-1}$, we either keep the summand the same (if $j=1$) or shrink the contribution of the negative terms if $i\in B$. Since \eqref{eq:3} is assumed to hold and this modification to the summands increases the positive contribution while decreasing the negative contribution of the summands, the inequality \eqref{eq:4} is holds for any positive integer $j$ as well.

If we sum \eqref{eq:4} over $j=1,\dots, k$ we get
$$
\sum_{i=1}^n (x_i ^k-1)  \frac{x_i^4+x_i^2+1}{x_i^2+x_i+1} \geq 0.
$$
Adding \eqref{eq:2} to this yields \begin{equation}\label{eq:example'}
\sum_{i=1}^n x_i^k \, \frac{x_i^4+x_i^2+1}{x_i^2+x_i+1} \geq n,
\end{equation} under the same assumptions of the original problem \eqref{eq:example}: $k, n\in \mathbb{Z}^{+}$, $x_i>0$ and $x_1+\cdots+x_n =n$. 
 
Finally, using inequality \eqref{eq:1} on the right-hand side of \eqref{eq:example'}, we prove \eqref{eq:example}. One highlight is that we proved these inequalities without any prior knowledge of any mathematical inequalities.  Moreover, note that \eqref{eq:example'} is a sharper inequality than \eqref{eq:example}.

\subsection{Implementation}

An implementation of this method has been completed by Kauers: the \texttt{ProveInequlity} procedure in the \texttt{SumCracker} Mathematica package \cite{Kauers2006} does the identification of the variables to be included and their shifts automatically, and ships the statement to be refuted directly to the CAD implementation of Mathematica. The proof of \eqref{eq:3}, in Mathematica, then turns into a single command:
\begin{center}
\begin{verbatim}
ProveInequality[SUM[(x[k]-1)(1+x[k]^2+(x[k])^4)/(1+x[k]+(x[k])^2), 
{k,1,n}]>=0,Using->{x[n]>0,SUM[x[k],{k,1,n}]==n},Free->{x},Variable->n]
\end{verbatim}
\end{center}
which terminates with an answer in milliseconds.

\subsection{Suitability for SMT}

A key-point to stress is that Gerhold--Kauers method actually generates and answers satisfiability problems, in the form \eqref{eq:exists_statement}, with all the existential quantifiers hidden but there. At each attempt,  the Gerhold-Kauers method checks the initial conditions and it looks to see if the refuted induction-step \eqref{eq:neg_induction} is unsatisfiable. Furthermore, any known information about the pieces of $\phi$ can be tagged alongside of \eqref{eq:neg_induction} and get fed to the CAD machinery to further restrict the search space and get the desired unsatisfiable answer. Their implementation simply used the CAD implementation of Mathematica to see in which regions \eqref{eq:neg_induction} can be satisfied. However, neither where this formulae is satisfied, nor the cylindrical structure of the decomposition to refute satisfiability, is essential to the problem.  Thus CAD could be safely replaced with any SMT solver that can tackle \verb+QF_NRA+ and may benefit from the incremental data-structures their internal machinery usually possess.

\section{Appearance and Handling of Rational Functions}
\label{Sec3}

In the automated proof sketches of \eqref{eq:2} and \eqref{eq:3}, we already saw the possibility of rational functions arising. The shifts of the variables $\hat{Z}$ and $\tilde{Z}$, which were used to simulate the sum and their shifts, introduced a rational function in the induction hypothesis clauses. In those examples above, the rational function could be simplified to a polynomial expression, but this is not true in general. We see that the satisfiability problems coming from this method naturally introduces rational functions. 

Rational functions inclusion and handling in satisfiability problems seems to be a somewhat sensitive topic in the SMT community and there are discussions about whether and how best to allow SMT-LIB to include rational functions in its language. While we leave that discussion for later, we will mention some possible pre-processing ways that can help us remedy the situation in a mathematically consistent way.

Assume that we are given a satisfiability problem where one of the clauses includes multivariate rational functions after simplifications. For example \[\frac{P(\mathbf{x})}{Q(\mathbf{x})}\ \sigma\ \frac{F(\mathbf{x})}{G(\mathbf{x})}\] where $P, Q, F, G \in \mathbb{Q}[\mathbf{x}]$\footnote{In this discussion, the rational field $\mathbb{Q}$ can be replaced by the reals $\mathbb{R}$, but here we restrict ourselves to stay within the limits of the SMT-LIB language.}, $\gcd(P,Q)=\gcd(F,G)=1$, and $\sigma \in \{>,<,\geq, \leq, =,\not=\}$. We can simplify this problem to 0, by subtraction followed by any simplifications which handle a problem of the form 
\begin{equation}
\label{eq:rat_func}
\frac{f(\mathbf{x})}{g(\mathbf{x})}{\color{gray}:=\frac{P(\mathbf{x})G(\mathbf{x}) - F(\mathbf{x})Q(\mathbf{x})}{Q(\mathbf{x})G(\mathbf{x})} }\ \sigma\ 0,
\end{equation} 
with $\gcd(f(\mathbf{x}),g(\mathbf{x}))=1.$

Handling rational functions in a mathematically consistent way is straightforward when the relation is an equation or an inequation. If $\sigma$ is $=$ or $\not=$ we can simplify \eqref{eq:rat_func} as 
\[
f(x)\ \sigma\ 0 \land g(x) \not= 0.
\] 
There are two equivalent formulations of \eqref{eq:rat_func} in the polynomial language when $\sigma$ is an inequality. One way is to avoid any sign considerations for the denominator polynomial $g(\mathbf{x})$ and multiply both sides of the relation \eqref{eq:rat_func} with its square. However, the poles of the original rational function should not be forgotten and be reflected in the outcome. This way the equivalent formulation of \eqref{eq:rat_func} is 
\begin{equation}
\label{eq:BDC}
f(\mathbf{x})g(\mathbf{x})\ \sigma\ 0 \land g(\mathbf{x})\not=0.
\end{equation} 
The disadvantage of this method is the likely rise in the degrees of the variables. When $f(\mathbf{x})$ and $g(\mathbf{x})$ are multiplied together some variables can get out of reach of the degree dependent QE techniques, such as virtual term substitutions \cite{Sturm2018}. 

Another possibility is to consider the sign of $g(\mathbf{x})$ and split the problem into two pieces driven by the guards $g(\mathbf{x})>0$ and $g(\mathbf{x})<0$. The statement we get using this approach is 
\begin{equation}
\label{eq:DDC}
( g(\mathbf{x})>0 \land f(\mathbf{x})\ \sigma\ 0) \lor ( g(\mathbf{x})<0 \land 0\ \sigma\ f(\mathbf{x})).
\end{equation}
Although this time the degrees of the variables stay lower, the size of the logical problem has grown. If the satisfiability problem starts with $n$ clauses including rational functions this problem would split it to a disjunction of $2^n$ statements. 

We suggest that the handling of rational functions be left to the SMT solvers. If users make this choice they may inadvertently disadvantage a solver. We elaborate on this later in \S\ref{ssec:denom}.

\section{Dataset and Benchmarking}
\label{Sec4}

\subsection{Dataset}
\label{ss:dataset}

We went through most examples given in \cite{GK05, Kauers2006, Kauers2005, KauersPillwein2010a} and equations \eqref{eq:1}, \eqref{eq:2}, and \eqref{eq:3} (the parts of the proof of \eqref{eq:example} which can be proven automatically) to describe them as non-linear arithmetic satisfiability problems in the SMT-LIB language, creating a dataset of 300 new SMT-LIB benchmarks. This was done by translating the original CAD calls of the \texttt{ProveInequality} procedure to SMT-LIB using the \texttt{SMTLIB} package in Maple \cite{Forrest2017}. This package already identifies the existence of a rational function in a clause and adds the denominator-is-nonzero clause to the problem. 

When problems involved a rational function in these calls then we also created two additional equivalent formulations of the problem, by clearing out the denominators in the basic way \eqref{eq:BDC} and in the disjunctive way \eqref{eq:DDC} as demonstrated above. The original examples with only polynomials and these two later polynomial-made examples were submitted in the call for new benchmarks for the 2022 SMT Competition\footnote{\url{https://smt-comp.github.io/2022/}}. 

In one group of our problems (the \texttt{SignPattern} problems from \cite{GK05}) the original problems contains a $\sqrt{5}$.  Mathematica's \texttt{CAD} implementation could handle these, but  algebraic numbers are not permitted within the definition of \verb+QF_NRA+ in general.  Thus we introduced the clauses $y^2 =5 \land y>0$ to bring the problem into \verb+QF_NRA+.  We note that the iteration of these clauses created high exponents for the pseudo-variable $y$s:  this was left for the solvers to handle.

\subsection{Solvers}

The SMT solvers used in this benchmarking are Z3 (v 4.8.8) \cite{BdMNW19} and Yices (v 2.6.4) \cite{Dutertre2014}, which both utilise the NLSAT algorithm \cite{JdM12} for \verb+QF_NRA+; and CVC5-Linux \cite{CVC5} (v 1.0.0) which uses the Cylindrical Algebraic Coverings algorithm for \verb+QF_NRA+ \cite{ADEK21}.  These three were selected as the strongest performers on \verb+QF_NRA+ in recent years.

We also evaluated some of the tools in Computer Algebra Systems, Maple and Mathematica:  the versions used are Maple 2022 and Mathematica 12.0.0.  In Maple, we used the \texttt{RegularChains:-QuantifierElimination} command \cite{CM16} to eliminate the calls in \eqref{eq:exists_statement} format. We also used the soon to be released Maple package \texttt{QuantifierElimination} \cite{Tonks2020}. The former utilises CAD constructed via triangular sets technology and the latter CAD with Lazard projection interlaced with cubic virtual term substitutions. In Mathematica, we used the \texttt{CAD} command \cite{Strzebonski2006}, as was used by Kauers' \texttt{ProveInequality} originally; the QE function \texttt{Resolve} which utilises also other QE methods such as virtual term substitution; and the meta-solver \texttt{Reduce} which makes use of Mathematica's other solving tools in addition.

Besides Maple's \texttt{RegularChains} implementation, all the other functions and solvers accepted inputs with rational functions.  

\subsection{Benchmarking Methodology}

In general we followed the methodology explained in  \cite{BDG17}.  All benchmarks were undertaken on a computer running openSUSE Leap 15.3 with 16GB of RAM and an Intel Xeon CPU E5-1650 v3 running at 3.50 GHz. All functions were given 20 minutes to attempt each of these problems. 

We display our results visually using survival plots.  To produce these we first solve each problem $q_i$, noting the time $t_i$ (up to our chosen threshold of 1200 seconds). Then for each solver we sort the $t_i$ into increasing order. discard the timed-out problems, and plot points $(k, \sum_{i=1}^k t_i)$. This approach does not guarantee that the same problems are returned with an answer in the chosen threshold from different implementations. However, for the cumulative problem set survival plots effectively encapsulate a lot of information about the success rate and the total time taken to solve for the successful answers.

\section{Analysis}
\label{Sec5}

\subsection{Overall performance}
\label{ssec:overall}

\begin{figure}[b]
\begin{center}
\includegraphics[scale=.3]{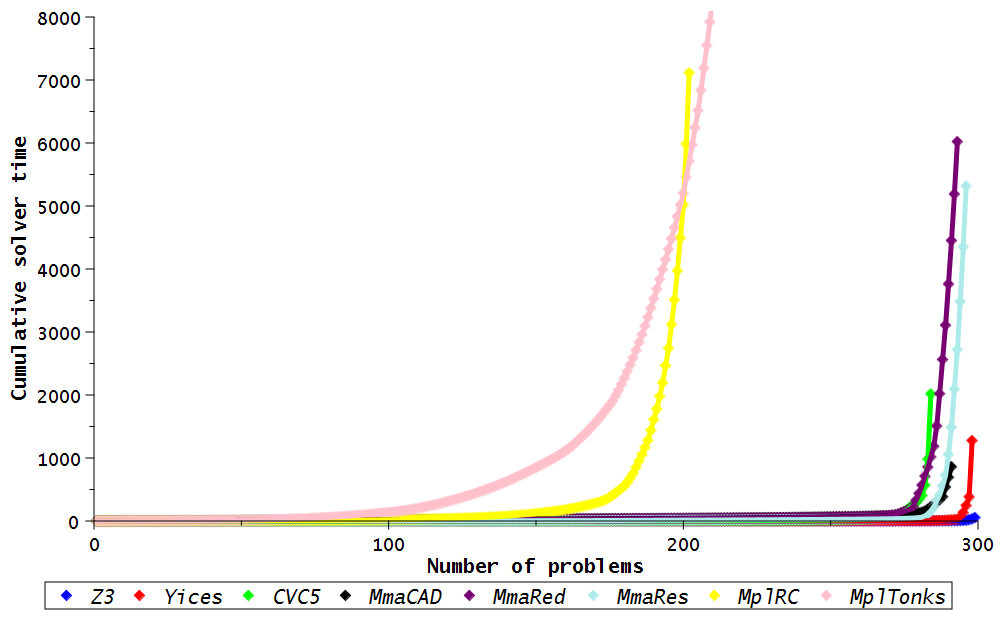}
\end{center}
\caption{Survival plot of benchmarks with the time scale up to 8000 seconds.}\label{Fig:SP1}
\end{figure}

\begin{figure}[t]
\begin{center}
\includegraphics[scale=.3]{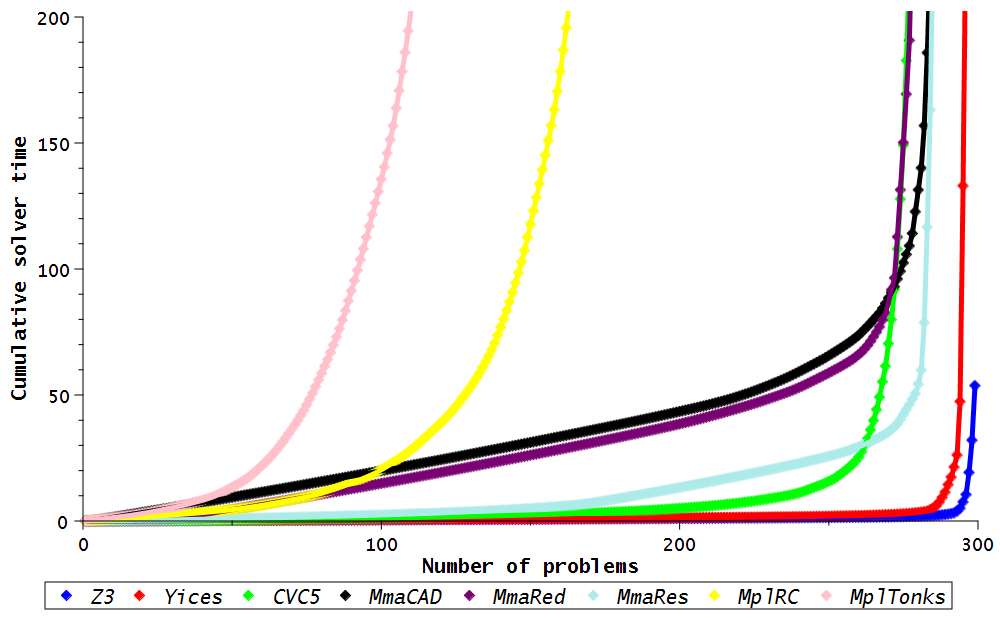}
\end{center}
\caption{Survival plot of benchmarks with the time scale up to 200 seconds.}\label{Fig:SP2}
\end{figure}

Figures \ref{Fig:SP1} and \ref{Fig:SP2} show the survival plots on different scales of the solver time.
It is clear that, for this dataset, \texttt{Z3} is superior: it timed-out only in one example.  It is then followed by \texttt{Yices} (timed out on 2 examples), then the various implementations in Mathematica (4, 7 and 9 time outs for \texttt{Resolve}, \texttt{Reduce} and \texttt{CAD} respectively) followed by \texttt{CVC5} (16 timeouts).  The two Maple functions performed far less well:  \texttt{RegularChains} did not accept rational functions at all, and both functions took far longer amounts of time to reach their conclusions.  We do note that Maple has also available direct calls to Z3 via \texttt{SMTLIB:-Satisfiable}.

It is not surprising that the SMT-solvers excel on satisfiability problems compared to full QE implementations using CAD.  Satisfiability is a sub-problem of QE with lower complexity.  What is surprising is that Mathematica's QE is competitive with the SMT-solvers.  The local projections used \cite{Strzebonski2016} may offer similar benefits to the model based SMT searches of \cite{JdM12}, \cite{ADEK21}; and we also note Mathematica has access to sophisticated logical simplification routines \cite{BS10}.  The DEWCAD project is working now to address the shortcoming's in Maple by building in Maple dedicated algorithms for satisfiability problems similar to those implemented in SMT \cite{BDESU21}.  


\subsection{Algebraic Number Substitutions}
\label{ss:algnum}

We suspect some timeout problems are due to a failure to substitute for algebraic numbers in the problems described in the final paragraph of Section \ref{ss:dataset}. In the \texttt{SignPattern} problems discussed there, the exponent of a variable $y$ (introduced to do bookkeeping of $\sqrt{5}$) in polynomials gets very high. The difficulty of this problem lowers immensely if a system can identify and at least utilize the second degree equational constraint $y^2=5$. We believe Z3 does this substitution and lowers the cost of calculations immensely. We also believe that most implementations would have been able to answer these questions in a matter of seconds if they were to do this preprocessing before asking for the satisfiability. For example, the CAD implementation of Mathematica can answer \verb+SignPattern\_Lemma4a-f+ examples from the dataset in about half a minute to a minute each, but when the $y^2=5$ is used and the degree of $y$ is reduced to a only linear powers, these numbers drop to under 10 seconds each.  We note that CVC5 performs particularly badly on these problems.

\subsection{Curiosities}
\label{ssec:curious}

On our examples, CVC5 is outperformed by Z3 and Yices overall, and it is outperformed by Mathematica for large compute times (see Figure \ref{Fig:SP1}). This is somehow at odds with the SMT Competition results of 2021\footnote{\url{https://smt-comp.github.io/2021/results.html}}. In addition to the algebraic numbers issue above, the poor performance may also be down to the presence of rational functions. Although CVC5 accepts rational functions in its input, we do not think that it does much preprocessing. The \texttt{SMTLIB} Maple package that was used to translate these examples to SMT-LIB language adds clauses to keep the denominators non-zero. Therefore, we never experience CVC5 encountering a division by zero and quitting or throwing an exception. However,  we observe that it takes a much longer time, and even times-out on occasion for the problems with rationals where others do not.

\begin{figure}[t]
	\begin{subfigure}{0.49\textwidth}
		\includegraphics[width=0.99\textwidth]{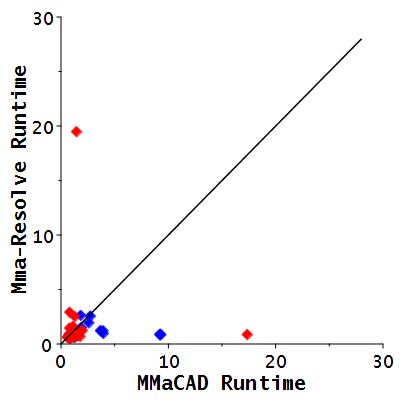}
		\caption{Matematica's \texttt{CAD} in comparison to its \texttt{Resolve}}
		\label{Fig:Mma}
	\end{subfigure}
	\begin{subfigure}{0.49\textwidth}
		\includegraphics[width=0.99\textwidth]{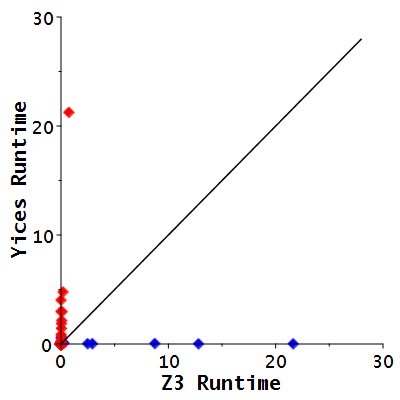}
		\caption{Z3 in comparison to Yices}
		\label{Fig:Z3Yices}
	\end{subfigure}
	\caption{Scatter plots of noticeable time differences.  Runtimes measured in seconds.  Blue data points are SAT examples and red are UNSAT.}
	\label{Fig:Scatter}
\end{figure}

Mathematica's \texttt{Resolve} and \texttt{Reduce} solve slightly more problems that its \texttt{CAD} procedure. But at one point \texttt{CAD} overtakes \texttt{Reduce}.  I.e. sometimes the cost of the extra considerations \texttt{Reduce} does hinders its success (see Figure~\ref{Fig:Mma}).  This indicates that there is scope for a better meta-algorithm to decide when \texttt{Reduce} resorts to \texttt{CAD}. 

Another curious observation between front runners Z3 and Yices is that Z3 is actually slower than Yices on the SAT problems. But since Z3 was much faster to identify that a problem is unsatisfiable and there were more UNSAT problems in the dataset, it gained victory overall. See Figure~\ref{Fig:Z3Yices}.  To the best of our knowledge Z3 and Yices both rely on NLSAT as the underlying theory algorithms. So it suggests the difference is in either the heuristics inside that, or the other incomplete methods tried first.

One can also observe from Figure~\ref{Fig:SP1} that \texttt{QuantifierElimination} Maple package is cumulatively slower than \texttt{RegularChains:-QuantifierElimination} on this problem set, overtaking eventually it only due to its handling of rational functions. Nevertheless, even when only considering examples with polynomial entries, \texttt{QuantifierElimination} is faster on 14\% of the examples. These examples might be where the virtual term substitution can be applied to make a significant difference.

\subsection{Effects of Denominator Clearance}
\label{ssec:denom}

Finally, we observe that \emph{how} we clear denominators affects conclusions about the best solver. We focus our attention to the rational function calls with their denominators cleared using \eqref{eq:BDC} or \eqref{eq:DDC}. For polynomial calls acquired by \eqref{eq:BDC}, Z3 is still the best solver but the second best solver changes hands from Yices to Mathematica's \texttt{Resolve} function both in time and in the number of problems solved. However, when we focus on rational call images under \eqref{eq:DDC} denominator clearance, we see that Mathematica \texttt{Resolve} solves one extra problem than Z3 and Yices. 

\section{Conclusions}
\label{Sec6}

Our first conclusion is that SMT solvers do very well on most of the examples in this problem set, outperforming computer algebra systems designed to tackle broader QE problems (Section \ref{ssec:overall}).  We also observe that the solvers perform differently on this new dataset than they did on the \verb+QF_NRA+ section of the SMT-LIB overall in the most recent competition. This shows us that it offers some new characteristics, and they continue the much needed diversification of the \verb+QF_NRA+ benchmarks. They also exposes some interesting strengths and weaknesses of solvers that the developers may find interesting to study (Section \ref{ssec:curious}).

Our second conclusion is that is a need for further work on the SMT-LIB language for \verb+QF_NRA+ to decide how best to deal with rational functions.  We observed that the choice of how we clear denominators effects conclusions over the best solver (Section \ref{ssec:denom}).  At the moment, the SMT-LIB seems to suggest the user should make this choice, but would it not be more appropriate for the solver to do it?  It clearly introduces a scope for new heuristics that researchers can explore.  The authors support that rational function calls be included in the SMT-LIB language, with a semantics that implies the denominator be non-zero.  But this must be defined so that the meaning is mathematically consistent and avoid getting conflicting results from solvers. 

Our third conclusion is in a similar vein: we suggest the SMT-LIB considers allowing the use of algebraic numbers in the input (Section \ref{ss:algnum}).  There are 21 examples under the \texttt{SignPattern} header, where we replaced $\sqrt{5}$ with a variable $y$ and two added clauses that $y^2=5 \land y>0$. Not only that, we let the iterations to grow the degree of $y$s and left the preprocessing to the solvers.  On this dataset the usually competitive CVC5 performed poorly.   But if we exclude this 21 problem subset, then among the polynomial calls CVC5 beats Mathematica methods in cumulative time. Allowing algebraic numbers in the problem statement stretches the definition of polynomial (usually assumed to have rational coefficients).  But many of the theory algorithms such as CAD can handle these and they can be encoded into actual polynomials.  Having the user do the encoding can make the problems artificially harder for solvers. 

\subsection*{Acknowledgements}

The authors would like to thank Manuel Kauers for providing a modified version of his \texttt{ProveInequality} function which exposed the CAD calls, easing the creation of the dataset.

All three authors are supported by the EPSRC DEWCAD Project (\emph{Pushing Back the Doubly-Exponential Wall of Cylindrical Algebraic Decomposition}): JHD and AU by grant number EP/T015713/1 and ME by grant number EP/T015748/1.  AU also acknowledges partial support from FWF grant P-34501N.

\bibliography{CAD}

\end{document}